\documentclass[epjST]{svjour}
\usepackage{graphics}
\usepackage{graphicx}
\usepackage{bm}
\usepackage{amssymb}
\usepackage{color}
\usepackage[usenames,dvipsnames]{xcolor}
\usepackage{subfigure}
\usepackage{amsmath}
\usepackage{amstext}
\usepackage{latexsym}
\usepackage[colorlinks=true,citecolor=Cerulean,linkcolor=Cerulean,urlcolor=Cerulean]{hyperref}
\usepackage{lipsum}

\usepackage{amsfonts}
\usepackage{epsfig}
\usepackage{mathrsfs}

\begin{document}
\title{Decay of a Thermofield-Double State in Chaotic Quantum Systems}
\subtitle{From Random Matrices to Spin Systems}
\author{A. del Campo\inst{1} \and J. Molina-Vilaplana\inst{2} \and L.~F. Santos\inst{3} \and J. Sonner\inst{4}}
\institute{Department of Physics, University of Massachusetts, Boston, MA 02125, USA 
\and 
Department of Systems Engineering, Technical University of Cartagena, C/Dr Fleming S/N. 30202, Cartagena, Spain
\and Department of Physics, Yeshiva University, New York, New York 10016, USA \and Department  of  Theoretical  Physics,  University  of  Geneva,
24  quai  Ernest-Ansermet,  1211  Gen\`eve 4,  Switzerland}
\abstract{
Scrambling in interacting quantum systems out of equilibrium is particularly effective in the chaotic regime. 
Under time evolution, initially localized information is  said to be scrambled as it spreads throughout the entire system. 
This spreading can be analyzed with the spectral form factor, which is defined in terms of the analytic continuation of the partition function. The latter is equivalent to the survival probability of a thermofield double state under unitary dynamics. Using random matrices from the Gaussian unitary ensemble (GUE) as Hamiltonians for the time evolution, we obtain exact analytical expressions at finite $N$ for the survival probability.  Numerical simulations of the survival probability with matrices taken from the Gaussian orthogonal ensemble (GOE)  are also provided. The GOE is more suitable for our comparison with numerical results obtained with a disordered spin chain with local interactions. Common features between the random matrix and the realistic disordered model in the chaotic regime are identified. The differences that emerge as the spin model approaches a many-body localized phase are also discussed.
 }

\maketitle
\section{Introduction}
\label{intro}

The study of information loss in black hole physics has  motivated the investigation of the rate at which quantum correlations can be scrambled~\cite{Hayden:2007cs,Sekino:2008he,Lashkari:2011yi,BR03,BR04,PR15}. In the context of the AdS/CFT correspondence, which connects gravity in a negatively curved spacetime (anti-de Sitter, AdS) with a strongly coupled field theory on the boundary of that spacetime (a conformal field theory, CFT) \cite{adscftbib,adscftbib2,adscftbib3,adscftbib4}, black holes are predicted to exhibit a very rapid mixing dynamics under unitary evolution \cite{Hayden:2007cs,Sekino:2008he}, approaching a bound on the exponential growth of out-of-time-ordered correlators~\cite{Maldacena15,SS15,Tsuji17}. The bounded quantity is analogous to a Lyapunov exponent in quantum chaotic systems, where properties of the energy spectrum are described by random matrix theory.

Within AdS/CFT, an eternal black hole is equivalent to a particular entangled state between two non-interacting copies of a CFT, a state referred to as the thermofield double (TFD) \cite{Maldacena:2001kr}. Scrambling is the redistribution of quantum information, initially localized in a subsystem, over the entire system under time evolution. More specifically, scrambling has been linked to certain out-of-time order four-point functions, as well as quantum information theoretic measures such as mutual information. 

It is natural to assess the spreading of information in a black hole via the survival probability of a TFD state, that is via the fidelity between the TFD state and its unitary time evolution~\cite{PR15,delcampo17}. This quantity is directly related to the analytic continuation of the partition function of the CFT~\cite{Dyer16,Cotler16,Cotler17} and ultimately, to the complex Fourier transform of the density of states. It is an extensively studied object in quantum chaotic systems~\cite{Leviandier1986,Guhr1990,Alhassid1992,Gorin2002,Torres2017Philo,TorresARXIV}.

The connection between black hole physics and random matrix theory has prompted  theoretical~\cite{Yao16,Swingle16,Laura17} and  experimental~\cite{GarttnerARXIV,Li17} efforts towards the investigation of information spreading in chaotic quantum systems in the laboratory. Realistic physical Hamiltonians generally are not described by full random matrices (FRM). FRM nevertheless successfully capture certain phenomena of realistic Hamiltonians \cite{MehtaBook,Brody1981}.  This motivates our search for properties that are common to FRM and realistic models with local interactions.

Recently a quantum many-body system of interacting fermions, the so-called Sachev-Ye-Kitaev (SYK) model ~\cite{kitaevKIPT,sadchevSYK}, has been proposed as a candidate for describing the physics of low dimensional quantum black-holes in holographic settings. The SYK model has quenched disorder with random all-to-all couplings, is maximally chaotic in the sense of \cite{Maldacena15,SS15} and connections with Random Matrix Theory \cite{Cotler16} and Eigenstate Thermalization Hypothesis \cite{Sonner17} have been established. The simplicity of the model has enabled several proposals for its study in the laboratory \cite{danshita16,Laura17}. Models that are yet closer to experimental systems include disordered spin chains, such as the one explored here.

In this work we focus on the characterization of information spreading via the dynamics of the survival probability of a TFD state. After identifying the salient features obtained with FRM, we compare the results with those of a one-dimensional (1D) spin-1/2 model with onsite disorder. This system is similar to the interacting fermionic system with quasi-random disorder experimentally implemented in~\cite{Schreiber2015}. We verify that the long-time dynamics of the FRM and of the spin model in the chaotic regime are very similar.

\section{Survival Probability of the Thermofield Double State}
\label{sec:1}
Consider a quantum system described by the Hamiltonian $H$ and a complete set of eigenstates $|n\rangle$, such that 
$H|n\rangle = E_n|n\rangle$. The spectral form factor $g(\beta,t)$, defined in terms of the analytic continuation of the partition function $Z(\beta,t)$, is written as
\begin{equation}
g(\beta,t) = |Z(\beta,t)|^2 = \sum_{n,m}\, e^{-\beta(E_n+E_m)-it(E_n-E_m)} ,
\end{equation}
where $\beta$ is the inverse temperature. The spectral form factor has been used~\cite{PR15,Maldacena15,Dyer16,Cotler16} to investigate the spectral properties of black holes and their information loss. Recently, $g(\beta,t)$ has been associated with the survival probability of a TFD state~\cite{delcampo17}. 

The TFD formalism for quantum systems is a strategy to treat a thermal mixed state, $\rho = e^{-\beta\, H}$, as a pure state in a bigger Hilbert space. This is done by considering two identical copies (labelled by $1$ and $2$) of the original quantum system. The degrees of freedom of these two copies are not coupled in any way. The states in the tensor product of the two system copies are $|m\rangle_1 |n\rangle_2$. The TFD state is a particular entangled state of the degrees of freedom of the two copies given by
\begin{equation}
\label{eq2.1}
| \Psi(\beta) \rangle = \frac{1}{\sqrt{Z(\beta)}}\, \sum_n\, e^{-\beta E_n/2}|n\rangle_1\, |n\rangle_2\,,
\end{equation}
from which it follows that $\rho =\rho_1 = {\rm Tr}_2\, | \Psi(\beta) \rangle\langle\Psi(\beta)| = e^{-\beta H}$.

The evolution of the TFD state under the Hamiltonian ${\cal H}=H_{1} \otimes \mathbb{I}_2$, where $H_1\equiv H$, is written as 
\begin{equation}
\label{eq2.2}
|\Psi(\beta,t)\rangle= e^{-it{\cal H}}|\Psi(\beta)\rangle=\frac{1}{\sqrt{Z(\beta)}}\sum_n e^{-(\beta/2+it)E_n}|n\rangle_1\, |n\rangle_2.
\end{equation}
The survival amplitude of the TFD state is then defined as the probability amplitude for the time-dependent state (\ref{eq2.2}) to be found in the initial state, i.e., the overlap
\begin{equation}
\label{eq2.3}
A(\beta,t)=\langle\Psi(\beta,0)|\Psi(\beta,t)\rangle=\frac{1}{Z(\beta)} \sum_n e^{-(\beta+it)E_n}. 
\end{equation}
The survival probability for this state is then 
\begin{equation}
\label{eq2.4}
S(\beta,t)=|A(\beta,t)|^2=\frac{1}{Z(\beta)^2} \sum_{n \neq m}\, e^{-\beta(E_n+E_m)-it(E_n-E_m)} + \frac{Z(2\beta)}{Z(\beta)^2},
\end{equation}
which, from definitions above, relates to the spectral form factor as
\begin{equation}
\label{eq2.5}
S(\beta,t)=\left|\frac{Z(\beta,t)}{Z(\beta)}\right|^2 = \frac{g(\beta,t)}{Z(\beta)^2}.
\end{equation}

This formulation allows to map unitarity constraints on the decay of general quantum systems, often expressed in terms of a survival probability (fidelity) decay, to the spectral properties of quantum systems. In Ref. \cite{delcampo17} bounds on the decay of the survival probability of the TFD at short times were derived that follow from unitarity constraints in conjunction with very general properties of the density of states (DOS). 
 
\subsection{General features of quantum decay}
\label{Sec:short}
Exploiting the power-series expansion of the time-evolution operator, it is straightforward to show that under unitary evolution, the initial decay of the survival probability is quadratic in time and controlled by the energy uncertainty $\Delta E_0$ of the initial state~\cite{Chiu77,FGR78},
\begin{equation}
\label{eq2.9}
S(\beta, t)=1-\left(\frac{t}{\tau_Z}\right)^2+\mathcal{O}(t^3), \hspace{0.7 cm} \tau_Z\equiv\frac{1}{\Delta E_0},
\end{equation}
where $\tau_Z$ is the Zeno time. The definition of the specific heat $c_V=k_B\beta^2 \Delta E_0^2$ for the TFD state~\cite{PR15,delcampo17}  leads to the equality $\tau_Z=\beta\sqrt{\frac{k_B}{c_V}}$.

Subsequently,
 the decay of the survival probability depends on the shape of the energy distribution of the initial state. This distribution is often referred to as local density of states (LDOS) and is given by
\begin{equation}
\rho_{\rm ldos} (E) = \sum_n  p_n \delta (E - E_n) ,
\label{eq:LDOS}
\end{equation} 
where $p_n$ is the absolute square of the overlap between the initial state and the energy eigenbasis.  When the initial state is a TFD state, $p_n=|\langle\Psi(\beta) |n\rangle_1\, |n\rangle_2 |^2 = e^{-\beta E_n}/Z(\beta)$. These components are the Boltzmann factors that define the occupation probability in the canonical thermal density matrix. They work as a filter function, enhancing the weight of the low energy eigenstates as $\beta$ increases.

The survival probability is the absolute square of the Fourier transform of the LDOS. For TFD states, $S(\beta,t)$ can also be written in terms of the density of states (DOS), $\rho(E)=\sum_nN_E\delta(E-E_n)$, with  $N_E$ being the degeneracy. We have
\begin{equation}
\label{eq2.12}
S(\beta,t) = \left| \int_{E_l}^{E_u} dE \rho_{\rm ldos} (E) e^{-iEt}  \right|^2 = \left| \frac{1}{Z(\beta)}\int_{E_l}^{E_u} dE\, \rho(E)e^{-(\beta +it)E} \right|^2,
\end{equation}
where $E_l$ and $E_u$ are the lower and upper bounds of the spectrum. 

The usual exponential decay of the survival probability is caused  by a Lorentzian LDOS. However, any other shape of $\rho_{\rm ldos} (E)$ necessarily leads to non-exponential decays~\cite{Khalfin57,Chiu77,Knight77}. Different behaviors have been characterized, ranging from power-law to superexponential decay \cite{delcampo17,Khalfin57,Chiu77,Knight77,Martorell09,Urbanowski09,delcampo11,Torres2014PRA,Torres2014NJP,Torres2014PRAb,Torres2014PRE,delcampo16}.

\subsection{Long-time power-law decay}
\label{sec.LongTimeDecay}
Due to the existence of a ground state $E_l$  in the energy spectrum, the long-time decay of the survival probability is characterized by a slower than exponential decay~\cite{FGR78,Khalfin57}. Mathematically, this is explained as follows. If the spectral density $\rho(E)$ vanishes for  $E<E_l$, the Paley-Wiener theorem implies that \cite{delcampo17}
\begin{equation}
\label{eq2.13}
S(\beta,t)\geq C\exp\left(-\gamma\, t^q\right) ,
\end{equation}
with $C,\gamma>0$ and $q<1$. In physical terms, the origin of this slower decay can be traced back to the possibility that the time-evolving state reconstructs the initial  state \cite{FGR78,Ersak69,Beau17}. 

The specific form of the long-time decay depends on the behavior of the LDOS near any edges. Under quite general conditions, a power-law decay holds. Indeed, if $\rho(E)\sim E^k$ near the edges $E_l$ and $E_u$, in the case above, the survival probability at long times is given by
\begin{equation}
\label{eq2.14}
S(\beta,t)\propto(t^2+\beta^2)^{-(k+1)}.
\end{equation}

\subsection{Long-time asymptotics: plateau}
For  systems with a continuum spectrum, the survival probability of an arbitrary initial state vanishes identically at $t\rightarrow \infty$ \cite{FK47}. For finite systems with a discrete spectrum and not too many degeneracies,  $S(\beta,t)$ eventually saturates to its infinite time average $\overline{S}(\beta)$, simply fluctuating around this value. 
Using Eq.~(\ref{eq2.4}), one sees that the value of this plateau for the TFD state is 
\begin{equation}
\label{eq2.11}
\overline{S}(\beta)=\lim_{T\rightarrow\infty}\frac{1}{T}\int_0^T dt \, S(\beta,t)=\frac{Z(2\beta)}{Z(\beta)^2}.
\end{equation}
This average is set by the purity of the canonical thermal density matrix, $\mathcal{P}(\rho_1)={\rm Tr}_1 \rho_1^2$. More generally, it coincides with the inverse participation ratio, ${\rm IPR}=\sum_{n}p_n^2$, which measures the  level of delocalization of the initial state in the energy eigenbasis, or equivalently, with $e^{-S^{(2)}}$ where $S^{(2)} = - \ln \left( {\rm Tr}_1\, \rho_1^2 \right)$ is the second R\'enyi entropy.  

\subsection{Correlation hole and ramp}
The behavior of the survival probability between the power-law decay and the plateau depends on the level of correlations between the eigenvalues. Probing the dynamics at these late times amounts to probing short- and long-range correlations between the eigenvalues. If the eigenvalues are correlated, as we find in chaotic systems, $S(\beta,t)$ reaches values below $\overline{S}(\beta)$. This dip is known as the `correlation hole' and was first explored in the context of FRM for initial states corresponding to random vectors~\cite{Leviandier1986,Guhr1990,Alhassid1992,Gorin2002}. The hole comprises the entire region where  $S(\beta,t)$ is below $\overline{S}(\beta)$, including the ramp to finally achieve the plateau. If the eigenvalues are uncorrelated, the correlation hole does not exist.

\section{Full Random Matrix Theory Models}
\label{sec:2}
We consider  a TFD state evolving under a FRM  taken from the Gaussian Unitary Ensemble (GUE). In the asymptotic limit of matrices with large rank $N$, the DOS obeys Wigner's semicircle law $\rho(E)=\frac{2N}{\pi \varepsilon}\sqrt{1-\left(\frac{E}{\varepsilon}\right)^2}$ in the interval $E\in[-\varepsilon,\varepsilon]$ and $\rho(E)=0$ everywhere else \cite{MehtaBook}. We choose random entries so that $\varepsilon=\sqrt{2N}$. The Fourier transform of the semicircle then leads to~\cite{delcampo17,Torres2014PRA}
\begin{equation}
S(\beta, t) = \frac{4 N^2 J_1 [\varepsilon (t + i \beta)] J_1 [\varepsilon (t - i \beta)]}{\varepsilon^2 (t^2 + \beta^2) Z(\beta)^2},
\label{Eq:Bessel}
\end{equation}
where $J_1 (x)$ is the Bessel function of the first kind.

For $t\ll 1/\varepsilon$, Eq.~(\ref{Eq:Bessel}) shows the quadratic decay mentioned in Sec.~\ref{Sec:short}. The asymptotic expansion, on the other hand, reveals a power-law behavior $\propto t^{-3}$ \cite{delcampo17,TorresKollmar2015,Tavora2016,Tavora2017}. When the temperature is high ($\beta $ small), $t^{-3}$ governs the envelope of  a set of decaying oscillations. These oscillations fade away as $\beta$ increases. In the light of the Paley-Wiener theorem, the cause of the power-law behavior is the compact support of Wigner's semicircle law. Near the edge of the spectrum, the DOS scales as $\rho(E)\sim \sqrt{E}$, which, following  Eq.~(\ref{eq2.14}), gives $S(\beta, t) \propto t^{-3}$. 

Equation~(\ref{Eq:Bessel}) does not explain the behavior of $S(\beta, t)$ beyond the power-law decay. By relying only on the Fourier transform of the semicircle, we neglect the fact that the spectrum is discrete. The semicircle is merely the envelope of the DOS; it does not take into account the internal structure of the distribution and the existence of correlations between the eigenvalues.  At long times, the effect of these correlations becomes relevant, causing the correlation hole, which is later followed by the saturation of the dynamics to $\overline{S}(\beta)$. We can explore these aspects analytically in the GUE and to this end we now present an exact expression for $S(\beta, t)$ in the GUE, valid for any $N$ and any $\beta$.

\subsection{Exact analytical expression for finite $N$}
All the distinct features of the survival probability associated with the aforementioned  time scales are captured by an exact analytical expression for $S(\beta, t)$ valid for arbitrary $N$ and  inverse temperature $\beta$ \cite{delcampo17}. The derivation uses the method of orthogonal polynomials and focuses on the spectral form factor, 
\begin{equation}
\label{eq3.1}
\langle g(\beta,t) \rangle=\langle Z(\nu) Z(\nu^*)\rangle =  \int dE_1 dE_2 \left\langle \rho^{(2)}(E_1,E_2) \right\rangle e^{-\nu E_1 - \nu^* E_2} + \langle Z(2\beta) \rangle ,
\end{equation}
where $\langle \cdot\rangle$ denotes the GUE ensemble average, $\nu = \beta + it$, and $\rho^{(2)}(E_1,E_2)$ is the GUE two-level correlation function. The latter can be decomposed~\cite{MehtaBook} into the two-level cluster function, $\rho_c^{(2)}(E_1,E_2)$, and the DOS, as $\left\langle \rho^{(2)}(E_1,E_2) \right\rangle=  \left\langle\rho_c^{(2)}(E_1,E_2) \right\rangle + \langle\rho(E_1)\rangle\langle\rho(E_2)\rangle$. 
The spectral form factor can now be written in terms of three terms, each dominating the dynamics at a different time scale \cite{delcampo17,Cotler16}, 
\begin{equation}
\label{eq3.3}
\langle g(\beta,t) \rangle=  |\langle Z(\beta,t)\rangle|^2 + g_c(\beta, t) + \langle Z(2\beta) \rangle .
\end{equation}
The initial decay of $\langle g(\beta,t) \rangle$ is controlled by the DOS through
\begin{equation}
\label{SurvAmpGUE}
\langle Z(\beta, t)\rangle= \int_{\mathbb{R}} dE e^{-\nu E} \langle\rho(E)\rangle .
\end{equation}
The DOS for any value of $N$ and averaged over the GUE is given by $\langle\rho(E)\rangle=\sum_{j=0}^{N-1}\varphi_j(E)^2$, where 
$\varphi_j(E)=\dfrac{1}{(2^jj!\sqrt{\pi})^{1/2}}e^{-\frac{E^2}{2}}H_j(E)$ are the harmonic oscillator eigenfunctions and $H_j(E)$ are Hermite polynomials~\cite{MehtaBook}. After some manipulations, one obtains
\begin{equation}
\label{eq3.4}
\langle Z(\beta,t)\rangle = e^{\frac{\nu^2}{4}}L_{N-1}^1\left( -\frac{\nu^2}{2} \right)\,,
\end{equation}
where $L_{N-1}^1(x)$ is a generalized Laguerre polynomial. Equation~(\ref{eq3.4}) leads to the quadratic decay of the survival probability at very short times and the power-law decay $\propto t^{-3}$ at longer times.

The second term in Eq.~(\ref{eq3.3}) takes over at yet longer times. It corresponds to the double complex Fourier transform of $\langle\rho_c^{(2)}(E_1,E_2) \rangle $ and was computed exactly in~\cite{delcampo17}, 
\begin{equation}
\label{eq3.5}
g_c(\beta, t) = e^{\frac{1}{4} \left(\nu^2 + \bar\nu^2  \right)}\sum_{j,k=0}^{N-1}\left( \frac{\nu}{\bar\nu} \right)^{k-j} \left|\frac{\Gamma(j+1)}{\Gamma(k+1)}L_{j}^{k-j}\left( -\frac{\nu^2}{2} \right)\right|^2\,.
\end{equation}
The third term in Eq.~(\ref{eq3.3}) is the infinite time average. These two terms together cause the dip of $S(\beta, t)$ below the plateau, which then rises approximately linearly (ramp)  up to $\bar{S}(\beta)$.

\begin{figure*}[t]
\centering
\includegraphics*[width=5.0 in]{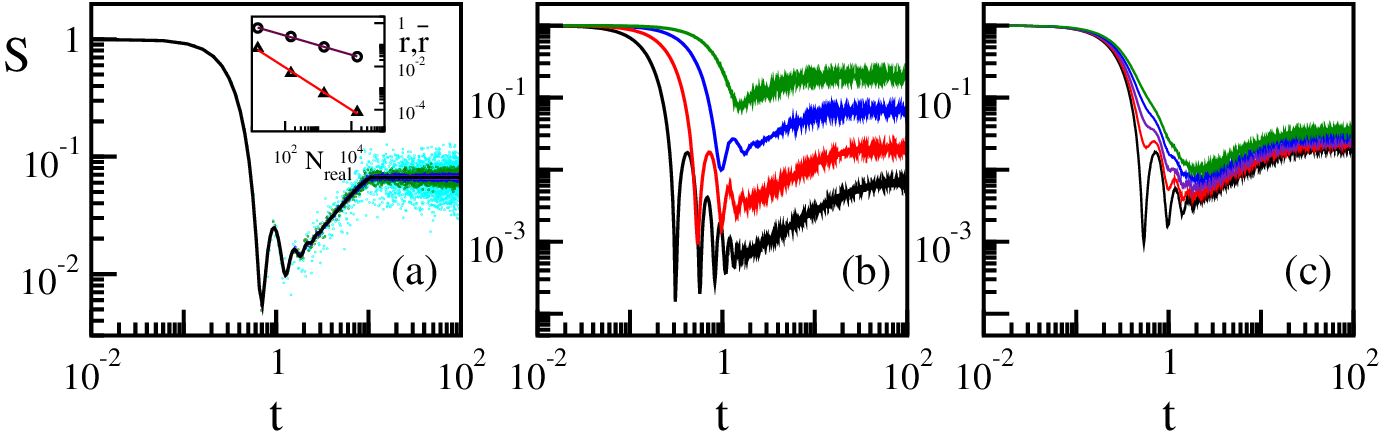}
\caption{(Color online) Survival probability for TFD states evolving under GUE (a) and GOE (b,c) FRM. In (a): Numerical vs analytical result, $\beta=0$, $N=15$, and data for $15,150,1500,15000$ realizations. As quantified in the inset, the numerical result uniformly converges to the exact result with the number of realizations: circles are for $r$ and triangles for $\bar{r}$. In (b): Numerical results for $\beta=0$ and different values of $N=5,15, 50,150$ (top to bottom). In (c): Numerical results for fixed $N=50$ and varying values of $\beta=0, 0.1, 0.15, 0.2, 0.25$ (bottom to top), each curve averaged over $1500$ realizations. }
\label{fig:GUEGOE}
\end{figure*}

Analytical results for the GUE at large $N$ have been obtained previously \cite{brezin1997spectral}, essentially via an expansion in $1/N$. Here, we emphasize that Eq.~(\ref{eq3.5}) is exactly valid  even for small or moderate values of $N$. Figure~\ref{fig:GUEGOE} (a) provides a comparison between numerical averages of samples drawn from the GUE and the analytical result for $N=15$. The convergence of the numerical results to the analytical curve as the number of realizations $N_{\rm real}$ increases is evident. A quantitative measure of this convergence can be assessed with the aid of
\begin{equation}
\label{Residuals}
r = \max_t\frac{ |S_{\rm exact}(t) - S_{\rm numerical}(t)|}{|S_{\rm exact}(t)|}
\end{equation}
i.e, through the maximum mismatch between the exact solution and the numerical one that can occur for all times. The inset of Fig.~\ref{fig:GUEGOE} (a) shows how this measure of residual goes to zero as a power of the number of realizations, $r \sim  N_{\rm real}^{-0.44}$, implying that indeed the numerics converge uniformly to the exact result. The inset also provides the residual measure $\bar{r}=|\overline{(S_{\rm exact}(t) - S_{\rm numerical}(t))/S_{\rm exact}(t)}|$, which converges to zero with a faster power law $\bar r \sim  N_{\rm real}^{-0.98}$. 

In Figs.~\ref{fig:GUEGOE} (b) and (c), we show numerical results for the survival probability using the Gaussian Orthogonal Ensemble (GOE). The outcomes are very similar to those for GUE, but the GOE is more suitable for our comparison with the spin model.  
In Fig.~ \ref{fig:GUEGOE} (b), $\beta=0$ and the curves are obtained for different values of $N$. The oscillations before the ramp fade away  as $N$ decreases. 
In Fig.~\ref{fig:GUEGOE} (c), we fix $N=50$ and let $\beta$ vary. The oscillations die out as $\beta$ increases. The gradual vanishing of the oscillations for smaller $N$ or lower temperature is a consequence of the reduced  number of levels participating on the dynamics.
The dependence of the oscillations on $N$ and temperature is observed also for the spin model, as we show next.

\section{Spin Systems}

Spin-1/2 models are frequently employed to describe systems investigated experimentally with magnetic compounds~\cite{Sologubenko2000,Hlubek2010}, nuclear magnetic resonance platforms~\cite{WeiARXIV}, ion traps~\cite{Jurcevic2014,Richerme2014}, and optical lattices~\cite{Trotzky2008}, among others. The particular disordered Hamiltonian that we use here has been extensively studied in the context of many-body localization~\cite{SantosEscobar2004,Nandkishore2015,LuitzARXIV} and is given by
\begin{equation}
H = \sum_{k=1}^{L} h_k S_k^z + J\sum_{k=1}^{L} \left( S_k^x S_{k+1}^x + S_k^y S_{k+1}^y + S_k^z S_{k+1}^z \right) ,
\label{ham}
\end{equation}
where $\hbar=1$, $L$ is the number of sites $k$, $S_k^{x,y,z}$ are the spin operators on each site, and the exchange interaction $J=1$. The amplitudes of the Zeeman splittings are random numbers $h_k$ from a uniform distribution in $[-h,h]$, where $h$ is the disorder strength.  Closed (periodic) boundary conditions are assumed. The Hamiltonian above commutes with the total spin in the $z$-direction, ${\cal S}^z=\sum_{k=1}^L S_k^z$. We assume that $L$ is even and use the subspace that has $L/2$ spins pointing up in $z$, so the dimension of the Hamiltonian matrix is $N=L!/(L/2)!$.

\subsection{Onset of chaos and LDOS}
When $h=0$, Hamiltonian (\ref{ham}) is integrable and solved with the Bethe ansatz~\cite{Bethe1931}. When $h$ is above a critical point $h_c$ (estimated to be $\sim 3.8$),  the system transitions to a spatially many-body localized phase. In these two extremes, $h=0$ and $h>h_c$, the eigenvalues are uncorrelated and can cross. The distribution $P$ of the spacings $s$ of neighboring unfolded levels is Poissonian, $P_P(s) = \exp(-s)$. 

\begin{figure*}[htb]
\centering
\includegraphics*[width=4. in]{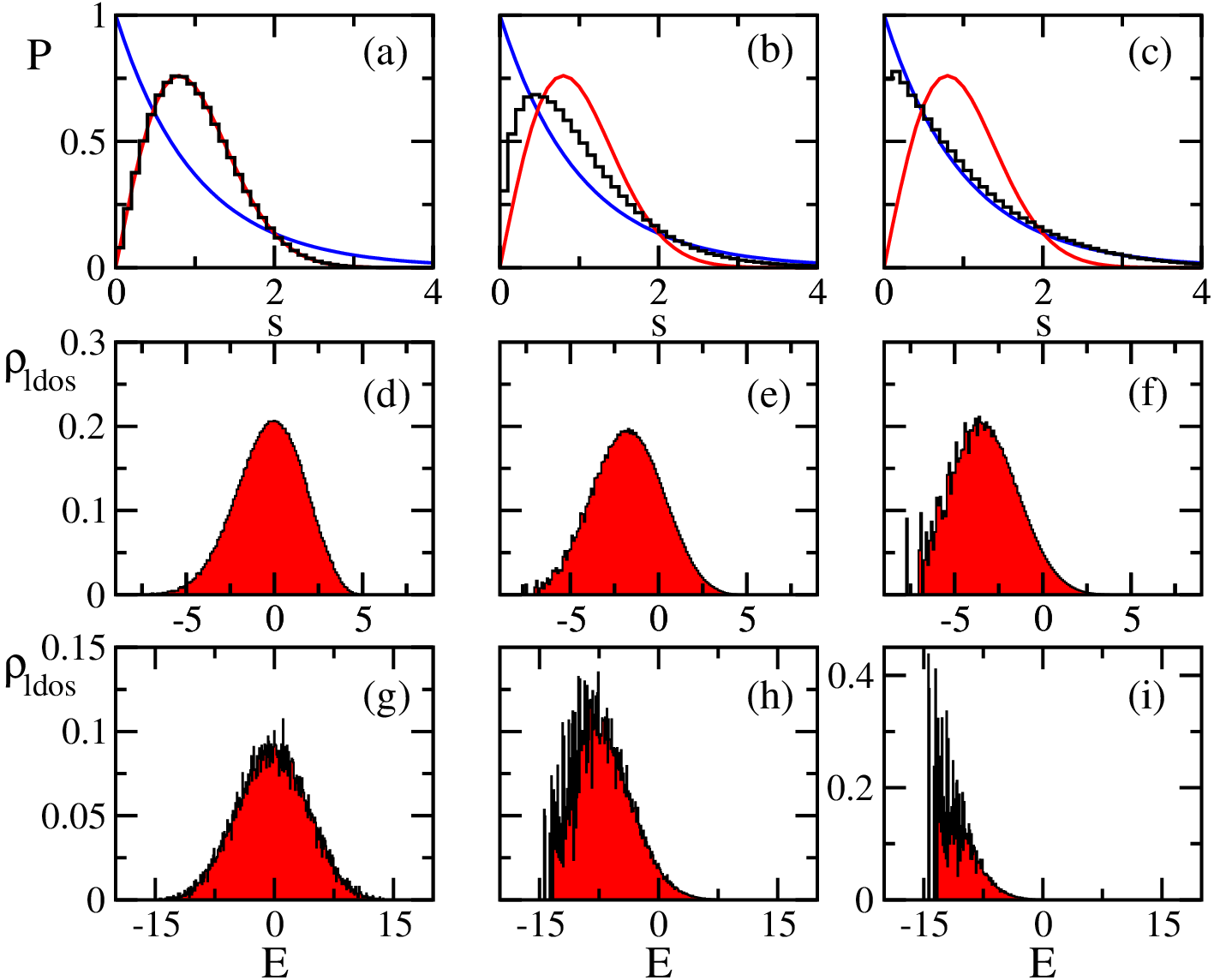}
\caption{(Color online) Level spacing distribution (a)-(c) and LDOS (d)-(i) for the spin model with different disorder strengths and TFD states at different temperatures; $L=16$. Disorder strengths: $h=0.5$ in (a), (d), (e), (f); $h=2.5$ in (b); $h=4.0$ in (c), (g), (h), (i). Temperatures: $\beta=0.0$ in (d) and (g); $\beta=0.4$ in (e) and (h); $\beta =0.8$ in (f) and (i). The level spacing distributions are averaged over 70 random realizations;  curves for the Poisson and Wigner-Dyson distribution are presented for comparison. The LDOS are shown for a single realization.}
\label{fig:PsLDOS}
\end{figure*}

In the region where $0<h<h_c$, the energy levels become correlated and repel each other~\cite{SantosEscobar2004,Santos2004,Dukesz2009}. For $L=16$, the strongest level repulsion occurs for $h\sim 0.5$ \cite{Torres2017}, where we find excellent agreement with the Wigner-Dyson distribution. Since the Hamiltonian matrix for Eq.~(\ref{ham}) is real and symmetric, the shape of this distribution is the same as that obtained for GOE FRM,  $P_{WD}(s)=\frac{\pi }{2} s \exp \left( {-\frac{\pi}{4} s^2} \right)$. 
Figures~\ref{fig:PsLDOS} (a), (b), and (c) illustrate the changes in the level spacing distribution as $h$ increases from 0.5 to $4.0$.

Contrary to FRM, where the DOS is semicircular, the DOS for the spin model has a Gaussian shape, as is typical of realistic systems with two-body interactions~\cite{Brody1981}. The form of the LDOS depends on both the DOS and the initial state. It becomes similar to the DOS when the initial state is highly delocalized in the energy eigenbasis~\cite{Torres2014PRA,Torres2014NJP}. In the case of TFD initial states, the LDOS coincides with the DOS  when $\beta=0$ [see Eq.~(\ref{eq:LDOS})]. 

In Fig.~\ref{fig:PsLDOS}, we show some examples of the shape of the LDOS for $h=0.5$ [(d)-(f)] and $h=4$ [(g)-(i)] for different values of the temperature (decreasing from the left to the right panels). The Gaussian shape for $\beta=0$  is evident in Fig.~\ref{fig:PsLDOS} (d) and Fig.~\ref{fig:PsLDOS} (g). As the temperature decreases  and the energy of the initial state, $E_0=  \sum_n p_n E_n$, is pushed closer to the low edge of the spectrum, the LDOS becomes asymmetric and more fragmented. The fragmentation is stronger for larger disorder strengths [compare Fig.~\ref{fig:PsLDOS} (f) and Fig.~\ref{fig:PsLDOS} (i)]. 

\subsection{Survival probability}

\begin{figure*}[ht]
\centering
\includegraphics*[width=4. in]{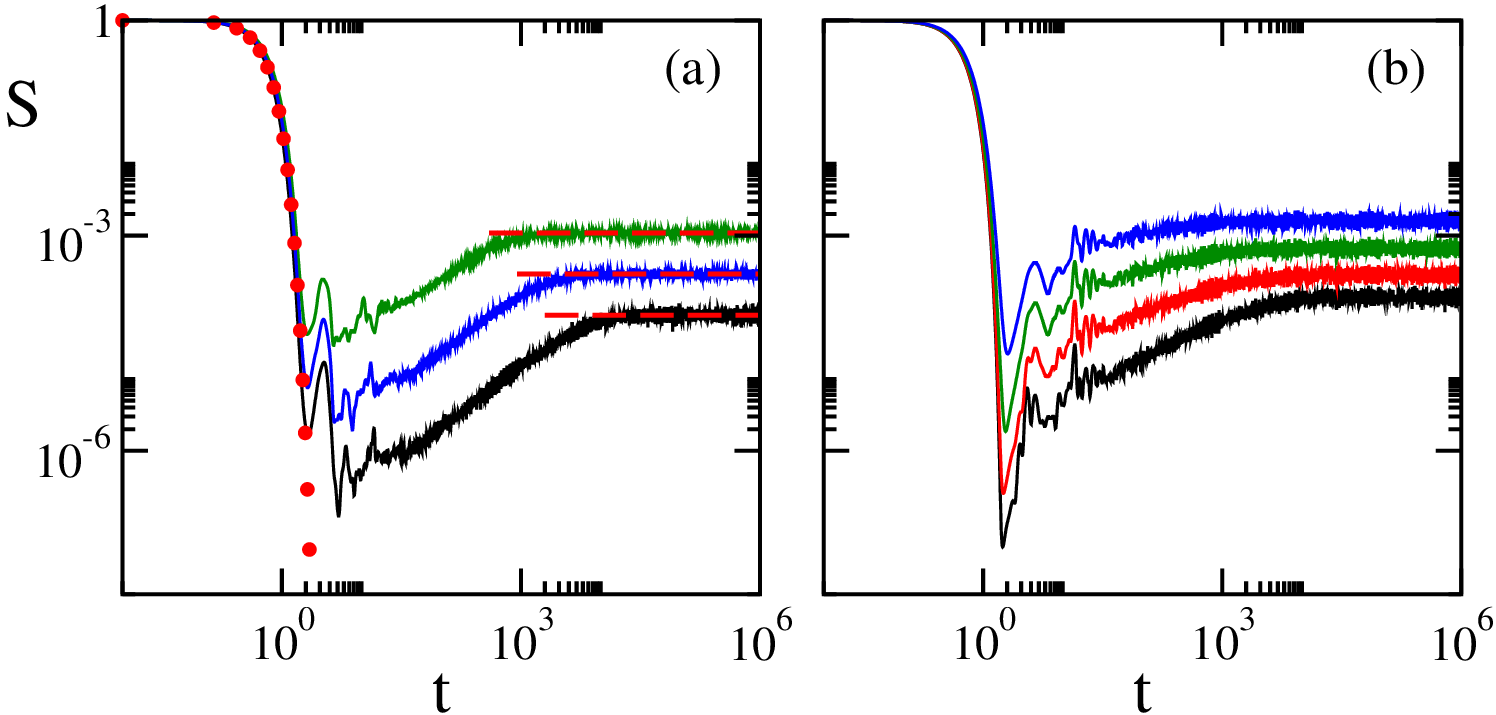}
\caption{(Color online) Survival probability for TFD states evolving under chaotic spin model with $h=0.5$. In (a): $\beta=0$ and the solid lines from bottom to top are for $L=16$ (average over 70 random realizations), $L=14$ (200 realizations), and $L=12$ (1000 realizations). The circles represent $\exp(-\Delta E_0^2 t^2)$.  The horizontal dashed lines correspond to $\overline{S} (\beta)$. In (b): $L=16$ and $\beta=0.4,0.6,0.8,1$ from bottom to top.}
\label{fig:hfixed}
\end{figure*}

The Fourier transform of a Gaussian LDOS leads to a Gaussian decay of the survival probability~\cite{Torres2014PRA,Torres2014NJP,Torres2014PRAb} that persists beyond the universal quadratic behavior of Eq.~(\ref{eq2.9}). We obtain $S(\beta,t) \propto \exp(- \Delta E_0^2 t^2)$, where $\Delta E_0=\sqrt{\sum_n p_n (E_n - E_0)^2} $
is the width of the LDOS. In Fig.~\ref{fig:hfixed} (a), we show the survival probability for the TFD state with $\beta=0$ evolving under the spin model with $h=0.5$. The circles indicate the Gaussian decay. The agreement with the numerics is excellent.

After the Gaussian decay, the curves in Fig.~\ref{fig:hfixed} (a) show oscillations that take us back to those seen for the GOE FRM in Fig.~\ref{fig:GUEGOE} (b). For the spin model, however, they take longer to develop and have smaller frequencies. For $L=16$, three oscillations are visible, while for $L=12$, the second one fades away. The disappearance of the oscillations as $N$ decreases also occurs for the FRM, which  suggests that more oscillations should become discernible for larger spin chains than those considered here.

As discussed in Sec.~\ref{sec:2},  the oscillations of the survival probability shows a power-law decay determined by the tails of the LDOS \footnote{This holds when the initial state is very much delocalized in the energy eigenbasis, as it happens for TFD states at high temperatures. In contrast, in studies of many-body localization, where the initial states are eigenstates of the Ising part of $H$ (\ref{ham}), as $h$ increases above 1, the LDOS becomes very sparse and the value of the power-law exponent gets smaller than 1. The dominating cause of the algebraic decay in this case is no longer the tails of the LDOS, but correlations between multifractal eigenstates~\cite{Torres2015,Tavora2016,Tavora2017}.}. For a Gaussian LDOS, when the ground state $E_l\ll \Delta E_0$, $\rho_{\rm{ldos}}(E)$ near $E_l$ is approximately constant. Using Eq.~(\ref{eq2.14}), we therefore expect $S(\beta, t) \propto t^{-2}$. This behavior is not clearly identified in Fig.~\ref{fig:hfixed} (a), but a hint of it can be noticed for initial states other than the TFD state (see~\cite{Tavora2016,Tavora2017}).

The oscillations in Fig.~\ref{fig:hfixed} (a) are followed by a ramp all the way to the saturation of the evolution at $\bar{S}(\beta) = \sum_n p_n^2$. This value is indicated with horizontal dashed lines. The function that describes the ramp is the same for the GOE FRM and for the chaotic spin model~\cite{TorresARXIV}, being given by the Fourier transform of $\rho_c^{(2)} (E_1,E_2)$ in Eq.~(\ref{eq3.3}).  At such long times, the evolution depends mostly on the level of correlations between the eigenvalues, which is similar for both models.

Figure~\ref{fig:hfixed} (b) shows the survival probability for the chaotic spin model for TFD states at different temperatures. Since deep in the chaotic region, the width of the LDOS does not change much as a function of $\beta$ [see Figs.~\ref{fig:PsLDOS} (d), (e), and (f)], the initial Gaussian decay is equivalent for all given curves. The plateau, on the other hand, naturally increases with $\beta$. Contrasting Fig.~\ref{fig:hfixed} (a) and Fig.~\ref{fig:hfixed} (b), we can make an analogy with the results in Fig.~\ref{fig:GUEGOE} (c) for FRM, where the oscillations are washed out as $\beta$ increases. For the spin model,  the oscillations observed for $\beta=0$ before the ramp [Fig.~\ref{fig:hfixed} (a)] are no longer seen already for $\beta=0.2$ [Fig.~\ref{fig:hfixed} (b)].

\begin{figure*}[t]
\centering
\includegraphics*[width=5.0 in]{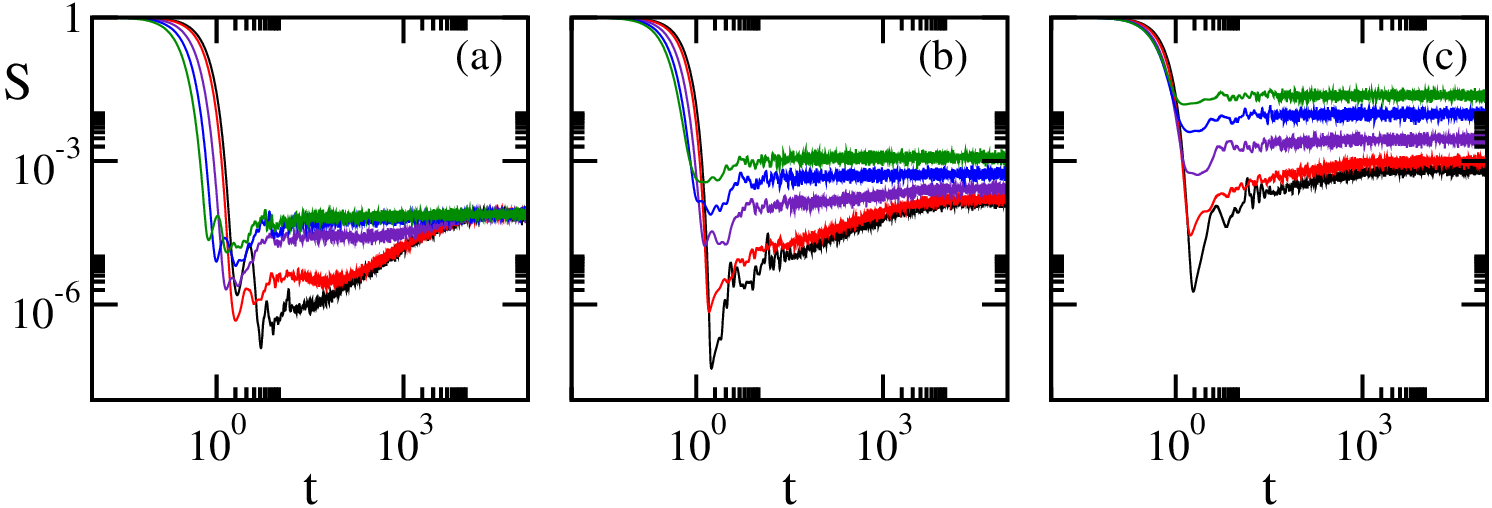}
\caption{(Color online) Survival probability for the disordered spin model. TFD states with $\beta=0.0$  (a), $\beta=0.4$  (b), and $\beta=0.8$ (c). Disorder strengths from bottom to top: $h=0.5, 1, 2, 3, 4$.  $L=16$; average over 70 random realizations. }
\label{fig:betafixed}
\end{figure*}

In Fig.~\ref{fig:betafixed}, we compare the numerical results for the spin model with different values of the disorder strength. The correlation hole gradually vanishes as $h$ increases and the correlations between the energy levels diminish [compare Fig.~\ref{fig:betafixed} with the level spacing distributions in Figs.~\ref{fig:PsLDOS} (a), (b), and (c)]. The correlation hole is a direct dynamical manifestation of the presence of level repulsion. It  disappears when the eigenvalues become uncorrelated. The depth of the hole can therefore be used as a signature of the integrable-chaos transition and, as a consequence, of the transition  to a many-body localized phase, as explored in Ref.~\cite{Torres2017Philo}. 

Contrasting Fig.~\ref{fig:betafixed} (b) and Fig.~\ref{fig:betafixed} (c), one also verifies that for the same $h$, the dip below $\bar{S}(\beta)$ becomes less pronounced for larger $\beta$. This occurs because for low temperatures, the TFD state predominantly samples the low-energy  part of the spectrum, where the energy levels are less correlated than those closer to the middle of the spectrum.

\section{Conclusions}
We have analyzed the scrambling of information of strongly interacting quantum systems via the survival probability of a thermofield double state. The comparison between the decay dynamics generated by FRM Hamiltonians and  by a realistic spin-1/2 model in the chaotic regime reveals common properties, which suggests that they are generic to chaotic many-body quantum systems out of equilibrium. 

The general properties identified include a very fast initial decay of the survival probability, a later power-law behavior determined by the tails of the LDOS, and the onset of a correlation hole that shows similar depth and an equivalent ramp up to the final plateau. For both models, we also see that the oscillatory behavior of the survival probability decreases as the temperature or the system size decreases.

We stress that the correlation hole, which is the main dynamical signature of quantum chaos, is not exclusive to the survival probability. It appears also in other experimentally observable quantities~\cite{TorresARXIV}, such as the density imbalance and out-of-time-ordered correlators \cite{GarttnerARXIV,Schreiber2015}. 

\begin{acknowledgement}
Acknowledgements. 
AdC acknowledges funding support from the John Templeton foundation and UMass Boston (project P20150000029279). LFS is supported by the NSF grant No. DMR-1603418 and thanks E.~J.~Torres-Herrera for several discussions about the correlation hole in spin models. JS is supported by the FNS grant No. 200021 162796 as well as the NCCR 51NF40-141869  The Mathematics of Physics (SwissMAP). JMV is supported by Ministerio de Econom\'{i}a y Competitividad FIS2015-69512-R and Programa de Excelencia de la Fundaci\'{o}n S\'{e}neca 19882/GERM/15.
\end{acknowledgement}


\begin{thebibliography}{99}

\bibitem{Hayden:2007cs}
P.~Hayden and J.~Preskill,
J. High Energy Phys. {\bf 0709}, 120 (2007).

\bibitem{Sekino:2008he}Y.~Sekino and L.~Susskind,
J. High Energy Phys. {\bf 0810},  065 (2008).

\bibitem{Lashkari:2011yi}
N.~Lashkari, D.~Stanford, M.~Hastings, T.~Osborne and P.~Hayden,
J. High Energy Phys. {\bf 1304}, 022 (2013).

\bibitem{BR03} J. L. F. Barbon and E. Rabinovici, J. High Energy Phys. {\bf 0311}, 047  (2003). 

\bibitem{BR04} J. L. F. Barbon and E. Rabinovici, Fortsch. Phys. {\bf 52},  642 (2004).

\bibitem{PR15} K. Papadodimas and S. Raju, Phys. Rev. Lett. {\bf 115}, 211601 (2015).

\bibitem{adscftbib} O. Aharony, S.~S. Gubser, J.M Maldacena, H. Ooguri and Y. Oz, Phys. Rep. {\bf 323},  183 (2000). 

\bibitem{adscftbib2} J. M. Maldacena, Adv. Theor. Math. Phys. {\bf 2}, 231 (1998). 

\bibitem{adscftbib3} S.~S. Gubser, I.~R. Klebanov, and A.~M. Polyakov, Phys. Lett. {\bf B428}, 105 (1998) 

\bibitem{adscftbib4} E. Witten, Adv. Theor. Math. Phys. {\bf 2}, 253 (1998) 
 

\bibitem{Maldacena15} J. Maldacena, S. H. Shenker, D. Stanford,  arXiv:1503.01409 (2015).

\bibitem{SS15} S. H. Shenker and D. Stanford, J. High Energy Phys. 05 (2015) 132.

\bibitem{Tsuji17} N. Tsuji, T. Shitara, M. Ueda, arXiv:1706.09160 (2017).

\bibitem{Maldacena:2001kr}
J.~M.~Maldacena, J. High Energy Phys. 04 (2003) 021.

\bibitem{delcampo17} A. del Campo, J. Molina-Vilaplana, J. Sonner, Phys. Rev. D {\bf 95}, 126008 (2017).

\bibitem{Dyer16} E. Dyer and G. Gur-Ari, arXiv:1611.04592 (2016).

\bibitem{Cotler16} J. S. Cotler, Guy Gur-Ari, M. Hanada, J. Polchinski, P. Saad, S. H. Shenker, D. Stanford, A. Streicher, M. Tezuka, arXiv:1611.04650 (2016).

\bibitem{Cotler17} J. Cotler, N. Hunter-Jones, J. Liu, B. Yoshida,  arXiv:1706.05400 (2017).

\bibitem{Leviandier1986} L. Leviandier, M. Lombardi, R. Jost,  J. P. Pique, 
Phys. Rev. Lett. {\bf 56}, 2449 (1986).

\bibitem{Guhr1990} T. Guhr and H. Weidenm\"uller, J. Chem. Phys. {\bf 146}, 21 1990).

\bibitem{Alhassid1992} Y. Alhassid, R.~D.Levine, Phys. Rev. A {\bf 46}, 4650 (1992).


\bibitem{Gorin2002}T. Gorin, T.~H. Seligman, Phys. Rev. E {\bf 65}, 026214 (2002).

\bibitem{Torres2017Philo}
E. J. Torres-Herrera and L. F. Santos, arXiv:1702.04363 (to appear at Phil. Trans. R. Soc. A , 2017).

\bibitem{TorresARXIV} E. J. Torres-Herrera, A. Garc\'ia-Garc\'ia, and L. F. Santos, arXiv:1704.06272.

\bibitem{Yao16} N. Y. Yao, F. Grusdt, B. Swingle, M. D. Lukin, D. M. Stamper-Kurn, J. E. Moore, E. A. Demler, arXiv:1607.01801 (2016).

\bibitem{Swingle16}
B. Swingle, G. Bentsen, M. Schleier-Smith, P. Hayden, Phys. Rev. A {\bf 94}, 040302 (2016).

\bibitem{Laura17}
L. Garc\'ia-\'Alvarez, I. L. Egusquiza, L. Lamata, A. del Campo, J. Sonner, E. Solano, Phys. Rev. Lett. {\bf 119}, 040501 (2017).

\bibitem{GarttnerARXIV} M. G\"arttner, J. G. Bohnet, A. Safavi-Naini, M. L. Wall, J. J. Bollinger, A. M. Rey, arXiv:1608.08938 (2016).

\bibitem{Li17}
J. Li, R. Fan, H. Wang, B. Ye, B. Zeng, H. Zhai, X. Peng, and J. Du, Phys. Rev. X {\bf 7}, 031011 (2017).

\bibitem{MehtaBook} M. L. Mehta, {\it Random Matrices.} (Elsevier, San Diego, 2004) Third edition.

\bibitem{Brody1981}
T. A. Brody, J. Flores, J. B. French, P. A. Mello, A. Pandey, and S. S. M. Wong. Rev. Mod. Phys. {\bf 53}, 385 (1981).


\bibitem{kitaevKIPT}  A. Kitaev,  Talks at KITP, April 7, (2015) and May 27, (2015).

\bibitem{sadchevSYK} S. Sachdev, Phys. Rev. X {\bf 5}, 041025, (2015).

\bibitem{Sonner17} J. Sonner, M. Vielma, arXiv:1707.08013 [hep-th]

\bibitem{danshita16} I. Danshita, M. Hanada, and M. Tezuka,  Prog. Theor. Exp. Phys. {\bf 8}, 083I01 (2017) .



\bibitem{Schreiber2015}
M. Schreiber, S. S. Hodgman, P. Bordia, H. P. L\"uschen, M. H. Fischer, R. Vosk, E. Altman, U. Schneider, and I. Bloch, Science {\bf 349}, 842 (2015).

\bibitem{Chiu77} C. B. Chiu, E. C. G. Sudarshan, and B. Misra
Phys. Rev. D {\bf 16}, 520 (1977).

\bibitem{FGR78} L. Fonda, G. C. Ghirardi, and A. Rimini,
Rep. Prog. Phys. {\bf 41}, 587(1978).

\bibitem{Khalfin57}  L. A. Khalfin, Sov. Phys. JETP {\bf 6}, 1053 (1958) [Zh. Eks. Teor.
Fiz. {\bf 33}, 1371 (1957)].

\bibitem{Knight77} P. L. Knight, Phys. Lett. A {\bf 61}, 25  (1977).

\bibitem{Martorell09} J. Martorell, J. G. Muga, D. W. L. Sprung, Lect. Notes Phys. {\bf 789}, 239 (2009).

\bibitem{Urbanowski09} K. Urbanowski,  Eur. Phys. J. D {\bf 54}, 25 (2009).

\bibitem{delcampo11} A. del Campo, Phys. Rev. A {\bf 84}, 012113 (2011).

\bibitem{Torres2014PRA} E. J. Torres-Herrera and L. F. Santos, Phys. Rev. A {\bf 89}, 043620 (2014).

\bibitem{Torres2014NJP}
E. J. Torres-Herrera, M. Vyas, and L. F. Santos, New J. Phys. {\bf 16}, 063010 (2014).

\bibitem{Torres2014PRAb} E. J. Torres-Herrera and L. F. Santos, Phys. Rev. A {\bf 90}, 033623 (2014).

\bibitem{Torres2014PRE}
E. J. Torres-Herrera and L. F. Santos, Phys. Rev. E {\bf 89}, 062110 (2014).

\bibitem{delcampo16} A. del Campo, New J. Phys. {\bf 18}, 015014 (2016).

\bibitem{Ersak69} I. Ersak, Sov. J. Nucl. Phys. {\bf 9}, 263 (1969).

\bibitem{Beau17} M. Beau, J. Kiukas, I. L. Egusquiza, A. del Campo,   Phys. Rev. Lett. {\bf 119}, 130401 (2017).

\bibitem{FK47} V.  A.  Fock and S. N.  Krylov,  Zh. Eksp. Teor. Fiz. {\bf 17}, 93 (1947).

\bibitem{TorresKollmar2015} E. J. Torres-Herrera, D. Kollmar, and L. F. Santos, Phys. Scr. T {\bf 165}, 014018 (2015).

\bibitem{Tavora2016}
M. T\'avora, E. J. Torres-Herrera, and L. F. Santos, Phys. Rev. A {\bf 94}, 041603 (2016).

\bibitem{Tavora2017}
M. T\'avora, E. J. Torres-Herrera, and L. F. Santos, Phys. Rev. A {\bf 95}, 013604 (2017).

\bibitem{brezin1997spectral}
E. Br\'ezin, S. Hikami,Phys. Rev. E {\bf 55}, 4067 (1997).

\bibitem{Sologubenko2000}
A. V. Sologubenko, K. Giann\'oo, H. R. Ott, U. Ammerahl, and A. Revcolevschi, Phys. Rev. Lett. {\bf 84}, 2714 (2000).

\bibitem{Hlubek2010}
N. Hlubek, P. Ribeiro, R. Saint-Martin, A. Revcolevschi, G. Roth, G. Behr, B. B\"uchner, and C. Hess, Phys. Rev. B {\bf 81}, 020405(R) (2010).

\bibitem{WeiARXIV}
K. X. Wei, C. Ramanathan, and P. Cappellaro, arXiv:1612.05249.

\bibitem{Jurcevic2014}
P. Jurcevic, B. P. Lanyon, P. Hauke, C. Hempel, P. Zoller, R. Blatt, and C. F. Roos, Nature {\bf 511}, 202 (2014).

\bibitem{Richerme2014}
P. Richerme, Z.-X. Gong, A. Lee, C. Senko, J. Smith, M. Foss-Feig, S. Michalakis, A. V. Gorshkov, and C. Monroe, Nature {\bf 511}, 198 (2014).

\bibitem{Trotzky2008}
S. Trotzky, P. Cheinet, S. F\"olling, M. Feld, U. Schnorrberger, A. M. Rey, A. Polkovnikov, E. A. Demler, M. D. Lukin, and I. Bloch, Science {\bf 319}, 295 (2008).

\bibitem{SantosEscobar2004} L. F. Santos, G. Rigolin, and C. O. Escobar, Phys. Rev. A {\bf 69}, 042304 (2004).

\bibitem{Nandkishore2015} R. Nandkishore and D. Huse, Annu. Rev. Condens. Matter Phys. {\bf 6}, 15 (2015).

\bibitem{LuitzARXIV} D. J. Luitz and Y. B. Lev, Ann. Phys.(Berlin) {\bf 529}, 1600350 (2017).



\bibitem{Bethe1931}
H. A. Bethe, Z. Phys. {\bf 71}, 205 (1931).

\bibitem{Santos2004}
L. F. Santos, J. Phys. A {\bf 37}, 4723 (2004).

\bibitem{Dukesz2009}
F. Dukesz, M. Zilbergerts, and L. F. Santos, New J. Phys. {\bf 11}, 043026 (2009).

\bibitem{Torres2017}
E. J. Torres-Herrera and L. F. Santos, Ann. Phys. (Berlin) {\bf 529}, 1600284 (2017).

\bibitem{Torres2015}
E. J. Torres-Herrera and L. F. Santos, Phys. Rev. B {\bf 92}, 014208  (2015).




\end{thebibliography}

\end{document}